\documentclass[apj]{emulateapj}
\shortauthors{Cardamone et al.}
\shorttitle{XR AGN}
\begin{document}
\title{Dust-Corrected Colors Reveal Bimodality in AGN Host Galaxy Colors at $z\sim1$.}
\author{Carolin N. Cardamone\altaffilmark{1,2}, C. Megan Urry\altaffilmark{1,2,3}, Kevin Schawinski \altaffilmark{2,3,4}, Ezequiel Treister \altaffilmark{4,5},  Gabriel Brammer\altaffilmark{1}, Eric Gawiser\altaffilmark{6}}
\email{carolin.cardamone@yale.edu}
\altaffiltext{1}{Department of Astronomy, Yale University, New Haven, CT 06511, USA}
\altaffiltext{2}{Yale Center for Astronomy and Astrophysics, Yale University, P.O.~Box 208120, New Haven, CT 06520}
\altaffiltext{3}{Physics Department, Yale University, PO Box 208120, New Haven CT 06520-8120}
\altaffiltext{4}{Einstein Fellow}
\altaffiltext{5}{Institute for Astronomy, 2680 Woodlawn Drive, University of Hawaii, Honolulu, HI 96822, U.S.A.}
\altaffiltext{6}{Department of Physics \& Astronomy, Rutgers University, Piscataway, NJ 08854-8019}

\begin{abstract}
Using new, highly accurate photometric redshifts from the MUSYC medium-band survey in the Extended Chandra Deep Field South (ECDF-S), we fit synthetic stellar population models to compare AGN host galaxies to inactive galaxies at $0.8 \le z \le 1.2$.  
We find that AGN host galaxies are predominantly massive galaxies on the red sequence and in the green valley of the color-mass diagram.  
Because both passive and dusty galaxies can appear red in optical colors, we use rest-frame near-infrared colors to separate passively evolving stellar populations from galaxies that are reddened by dust.
As with the overall galaxy population, $\sim$25\% of the `red' AGN host galaxies and $\sim$75\% of the `green' AGN host galaxies have colors consistent with young stellar populations reddened by dust.
The dust-corrected rest-frame optical colors are the blue colors of star-forming galaxies, which implies that these AGN hosts are not passively aging to the red sequence.
At $z\sim1$, AGN activity is roughly evenly split between two modes of black hole growth: the first in passively evolving host galaxies, which may be heating up the galaxy's gas and preventing future episodes of star formation, and the second in dust-reddened  young galaxies, which may be ionizing the galaxy's interstellar medium and shutting down star formation. 
\end{abstract}

\section{Introduction}
\label{intro}

The colors and morphologies of galaxies form a bimodal population at $0\la z\la2$ \citep[][]{blantonetal2003,belletal2004,borcheetal2006,deluciaetal2007,Ilbertetal2009mass,williamsetal2009,brammeretal2009}.
A red sequence contains massive galaxies and a blue cloud is composed of lower mass  galaxies, while relatively few galaxies lie in the  green valley in between.
The observed color distribution is consistent with galaxy color evolving from blue to red by the rapid quenching of star formation \citep{faberetal2007}.
A central AGN, which can heat up the galaxy's gas \citep{silkrees1998,dimatteoetal2008} and consequently shut down star formation, can cause this rapid color evolution \citep{faberetal2007,schawinskietal2007}.
Moreover, the observed correlation between the mass of the central black hole and the host galaxy's mass \citep[e.g.,][]{tremaineetal2002}
suggests a potential common cause of star formation and AGN accretion, e.g. mergers  \citep{hopkinsetal2006, cattaneoetal2009}.
Additionally, some models invoke subsequent AGN activity to maintain the passive nature of red galaxies,  by preventing future episodes of gas cooling \citep[e.g.,][]{crotonetal2006,boweretal2006}.

Several studies show that AGN host galaxies have intermediate host galaxy colors at $ z \la 1$, placing them in the green valley \citep{nandraetal2007,bundyetal2008, georgakakisetal2008,silvermanetal2008,schawinskietal2009agn}.
This observation is sometimes interpreted as evidence for AGN involvement in the transition of host galaxies from the blue cloud to red sequence.
However, if this were true, we should observe at least some AGN in blue galaxies, because it takes several hundred Myrs for a stellar population to age to the colors of the green valley \citep{schawinskietal2009agn}.  
Accordingly, if the intermediate colors of the AGN host galaxies are due to aging stars, AGN activity likely follows the shutdown of star formation with a delay of several 100 Myr.

Recent studies of galaxy colors suggest that the green valley may not consist solely of a transition population of galaxies with intermediate ages, but may contain galaxies from the blue cloud reddened by obscuring dust \citep{belletal2005,cowiebarger2008,brammeretal2009}.
The same could be true for AGN host galaxies, complicating the observational picture of the role of AGN feedback in galaxy color evolution.  
In this Letter, we use highly accurate photometric redshifts in the Extended {\it Chandra} Deep Field-South (ECDF-S, Cardamone et al.\ 2010) to investigate the host galaxies of X-ray selected AGNs at $0.8\le z\le1.2$.
Throughout this paper, we assume $H_0 = 71  \: {\rm km \: s^{-1} \: Mpc^{-1}}$, $\Omega_{\rm m} = 0.3$ and $\Omega_{\rm \Lambda} = 0.7$.

\section{Colors and Masses of Galaxies and AGN Hosts}
\label{data}
The combination of depth ($R_{AB}\sim26$), area (30\arcmin $\times$ 30\arcmin) and multi-wavelength coverage make the ECDF-S ideal for the study of moderate luminosity AGN and galaxies at $z\sim1$.
The MUSYC survey includes the optical ($U38UBVRIz$), near-  and mid-infrared ($JHK$ \& $3.6-8.0\mu$m) data \citep[][Damen et al.\ 2010]{gawiseretal2006,tayloretal2009} needed to investigate galaxy stellar populations and medium-band  observations that densely sample the galaxy Spectral Energy Distributions (SEDs), allowing for extremely accurate photometric redshift determinations (Cardamone et al.\ 2010).
Photometric redshifts were computed with EAzY \citep{brammeretal2008} using the full 32-band SEDs, which results in median[$\Delta z/ (1+z)] \la 0.01$ at $z\le 1.2$ (Cardamone et al.\ 2010).  
Here, we study galaxies with reliable photometric redshifts ($Q_z \le1$; \citealt{brammeretal2009}) at $0.8 \le z \le 1.2$.

We use deep X-ray imaging to identify galaxies with AGN activity since X-rays reveal all but the most obscured AGN \citep{branthasinger2005}.   
To remove potential interlopers whose X-ray emission comes primarily from star formation, we consider X-ray sources sources with ${\rm log (L_{XR}) \ge 42 \rm{~ergs~s^{-1}}}$.
We combine the deep 2 Ms imaging in the central $Chandra$ Deep Field South \citep{luoetal2008} with the shallower 250 ks observations of the ECDF-S flanking this field \citep{lehmeretal2005,viranietal2006}\footnote{The ECDF-S depth is sufficient to detect AGN with observed ${\rm log (L_{XR}) \ga 42 \rm{~ergs~s^{-1}}}$ out to $z\sim1.2$.}
Obscured AGN are identified by their observed X-ray hardness ratio, $HR=\frac{H-S}{H+S} \ge -0.2$, where H and S are the X-ray counts in the hard (2-8 keV) and soft (0.5-2 keV) bands.  
Of 1134 X-ray sources detected, 825 have reliable optical counterparts identified by a likelihood ratio method \citep{cardamoneetal2008}.
Photometric redshifts and rest-frame fluxes are computed following the procedure of \citet{brammeretal2009}.
Details of the redshift and rest-frame flux determinations are discussed in \citet{cardamoneetal2010}.
Combining the spectroscopic and photometric redshifts, there are 114 sources with $0.8\le z \le 1.2$ and  ${\rm log (L_{XR}) \ge 42 \rm{~ergs~s^{-1}}}$, of which 80 are obscured AGN ($HR \ge -0.2$).
We use this sample of 114 AGN to study host galaxy properties.  
At the modest luminosities sampled by the ECDF-S volume (only 8 sources have ${\rm log (L_{XR}) \ga 44 \rm{~ergs~s^{-1}}}$), most of the observed optical/IR light actually arises from the galaxy  \citep[e.g.,][]{cardamoneetal2007,bundyetal2008,silvermanetal2008}.
We revisit the relative contribution of AGN and star light in Section \ref{sec:dusty}.

We fit the galaxy and AGN host SEDs with single-burst stellar population models, for which the medium-band photometry is particularly useful. 
Using reliable spectroscopic redshifts where available (28 sources) and photometric redshifts otherwise (86 sources), 
 we employ FAST \citep{krieketal2009} to estimate the stellar mass, star-formation time scale, star-formation rate and dust content ($A_V$) for each galaxy.
We use stellar templates from \citet{maraston2005}, adopting the  \citet{kroupa2001} IMF and solar metallicity, and fitting \citet{calzettietal2000} dust extinction ($A_V$=0-4).
We obtain mass estimates and measurements of $A_V$ for all galaxies and the hosts of  X-ray-selected AGN from the SED fit, and run Monte Carlo simulations varying the photometry for estimates of uncertainty in the values.
Further uncertainties come from systematic effects due to assumptions of metallicity, galactic extinction laws and the adopted stellar templates, and can affect mass measurements by 0.1-0.6 dex
\citep[see e.g.,][]{muzzinetal2009, conroygunn2010}.
Degeneracies in the fitting are also a concern (e.g., the age--metallicity degeneracy), so we use only the measurements of stellar mass and dust content in this paper.
Using simulated galaxy SEDs we found adding light from a quasar template increases the fitted mass by 0.1 dex but had very little effect on the fitted dust content in models with ages of less than 1 Gyr ($A_{V(AGN+gal)}\sim A_{V(gal)}+0.1$) and no effect on the fitted $A_V$ for models with ages greater than 1 Gyr.

\begin{figure*} [ht] 
\includegraphics[angle=0, width=0.97\textwidth]{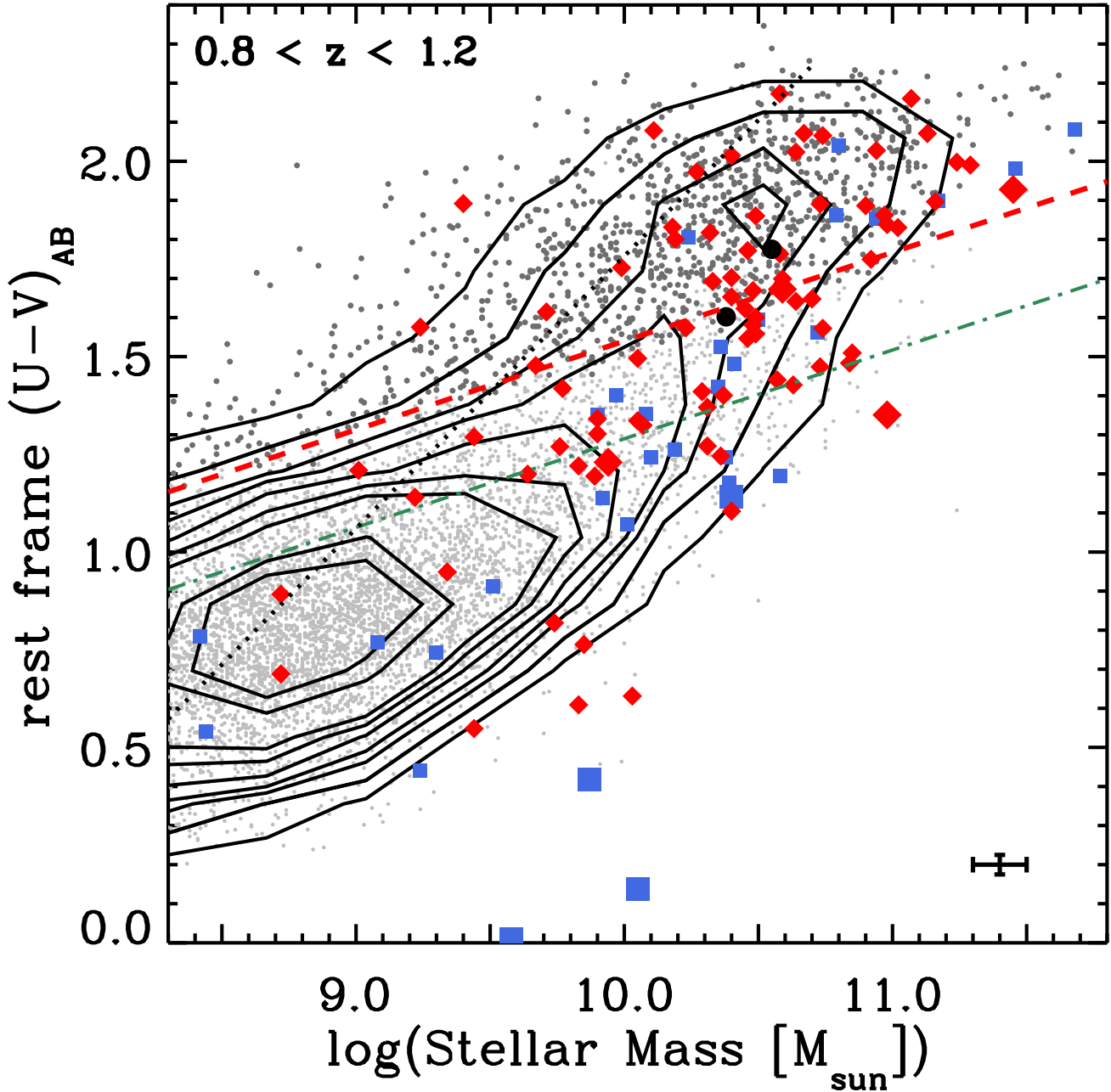}
\caption{The galaxy population (grey points, black contours; black dotted line indicating 90\% completeness) is bimodal in color verses stellar mass.
The red dashed lines show the red sequence cut defined by \citet{borcheetal2006} and the green dash-dotted line defines an approximate green valley region.  In contrast, the X-ray selected AGN (blue squares for unobscured AGN, $HR \le -0.2$;  red diamonds for the obscured AGN, $HR \ge -0.2$; and black circles for sources detected only in the full X-ray band) are distributed continuously, peaking at intermediate colors.
The luminous X-ray sources (${\rm log (L_{X}) \ge 44 \rm{~ergs~s^{-1}}}$; larger symbols) are likely biased by QSO light towards bluer colors and larger masses.    
Monte Carlo simulation were run in EAzY to address errors due to photometric uncertainties. Typical error bars are shown in the bottom right corner.
The presence of AGN host galaxies in the green valley has been interpreted to mean that AGN hosts are aging from the blue cloud to the red sequence.  However, these colors can also result from dust reddening.}
\label{fig:cmass}
\end{figure*}

The rest-frame $U-V$ color measures the strength of the 4000\AA~Balmer break in the galaxy spectrum and separates red galaxies dominated by older stellar populations, which have a strong Balmer break, from blue galaxies experiencing ongoing star formation.
In Figure \ref{fig:cmass}, we show $U-V$ color verses stellar mass for the galaxy population. 
We confirm the bimodal distribution of galaxy colors at $z\sim1$ \citep[e.g.,][]{belletal2004, Ilbertetal2009mass}.
In contrast,  the distribution of AGN host galaxies is continuous, extending from red to blue, as was found previously in color verses magnitude diagrams \citep[e.g.,][]{nandraetal2007,silvermanetal2008,treisteretal2009}.
We also confirm that AGN are preferentially in massive galaxies (${\rm M_{stellar} \ga 10^{10} M_{\odot}}$; \citealt[e.g.][]{brusaetal2009}).

The observed rest-frame colors of AGN host galaxies have been interpreted as indicating intermediate stellar population ages, supporting the idea that AGN activity plays a role in quenching star formation \citep[e.g.,][]{nandraetal2007}.
Indeed, at $z\sim 0$, AGN are in post-starburst galaxies as assessed via spectral analysis, not just color \citep{schawinskietal2007} and post-starburst galaxies show evidence of galactic winds potentially driven by AGN feedback \citep{tremontietal2007}.
However, observed colors are also affected by dust obscuration.
To break this degeneracy, multiple rest-frame colors can be used to separate quiescent galaxies from those that are actively star-forming but enshrouded by dust \citep[e.g.,][]{labbeetal2005,wuytsetal2007}.
We apply this technique to investigate the nature of the AGN host galaxy population.

\section{Passive \lowercase{vs} Dusty Galaxies}
\label{sec:dusty}
\begin{figure*}[ht]
\includegraphics[angle=0, width=0.97\textwidth]{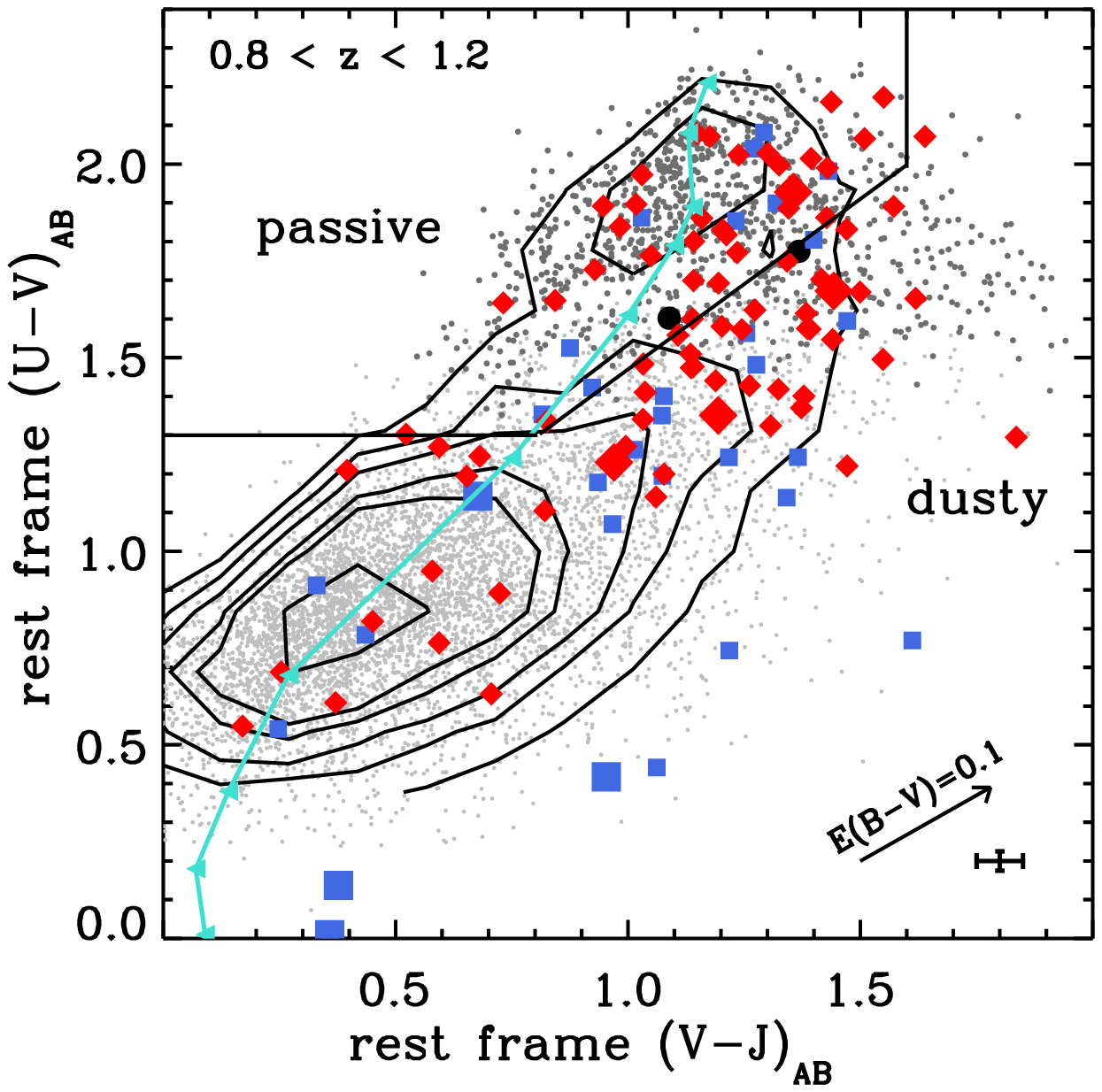}
\caption{ The distribution of galaxies in $U-V$ color verses $V-J$ color has two peaks, one at red colors inside the passive selection region and another at blue colors   (points as in Figure \ref{fig:cmass}).
Only half of the AGN host galaxies are located in the passive region of this diagram; the rest have colors indicative of star formation, primarily dust-obscured star formation.
For objects on the red sequence in Figure \ref{fig:cmass} (dark grey points), 20\% of galaxies and 20\% of AGN host galaxies lie in the dust-obscured region, i.e., outside the ``passive'' box.
Additionally, most ($\sim75 \%$) galaxies and AGN in the green valley are dust-obscured, rather than a transition population aging onto the red sequence.} 
\label{fig:col}
\end{figure*}

Both dusty galaxies and older stellar populations have red optical colors (e.g., $U-V$).
However, near-infrared light can distinguish between the two populations: given a similar 0.5 $\mu$m flux, a dusty population has more emission near $\sim 1~ \mu$m than an older stellar population.
Figure 2 shows  the rest-frame $U-V$ verses $V-J$ colors for galaxies in the ECDF-S, with the region occupied by passive galaxies indicated by a solid black line (red $U-V$ and blue $V-J$; \citealt{labbeetal2005,williamsetal2009}).  
The cyan line shows a synthetic stellar population composed of a young starburst ($\tau \sim100$ Myrs) and a more massive older stellar population (triangles indicate the model at ages of 0.025 - 5 Gyrs).
This synthetic population has the color of galaxies in the blue cloud for the first $\sim$300 Myrs, enters the passive box after $\sim400$Myrs, and joins the peak of red-sequence galaxies inside the  passive galaxy selection box 1 Gyr post starburst.
However, the model does not fall in the dusty region of the diagram unless significant extinction is added.

While the red-sequence galaxies (Figure \ref{fig:cmass}, dark grey points) cluster inside the passive galaxy region, $\sim$20\% of  red sequence galaxies (with log$M_{star}\ge 10$, a completeness cut for $0.8\le z\le1.2$) scatter towards redder $V-J$ colors, indicating the presence of dust.
This is consistent with previous findings at this redshift range \citep[e.g.,][]{cardamoneetal2010,brammeretal2009}.
Additionally,  $\sim$75\% of the green valley galaxies are consistent with dust-reddened colors, indicating that 
there are far fewer objects genuinely transitioning between the blue cloud and red sequence.
Similarly, $\sim$75\% of green valley and $\sim20\%$ of red sequence AGN host galaxies fall outside the passive selection region.  
This observation calls into question the interpretation of intermediate rest-frame $U-V$ colors of AGN host galaxies as evidence that AGN preferentially lie in galaxies aging from a blue star-forming phase to a red quiescent phase.

Overall, half of the AGN host galaxies are composed of passive stellar populations, the rest have colors indicative of younger stellar populations with varying amounts of dust reddening.
Of the AGN host galaxies (log$M_{star}\ge 10$), $\sim$50\% (38 out of 80) fall inside the passive selection region. The same fraction of obscured AGN host galaxies, which have minimal AGN contamination to their colors, lie inside the passive selection region (28 out of 57).

At $z\sim1$, it is difficult to identify light from a central nuclear source, but with HST resolution it is possible  to subtract the central point source and investigate uncontaminated host-galaxy colors.    
We compare the point source-subtracted host galaxy color to the total integrated color for the AGN in our sample imaged by {\it HST} in the GOODS survey (Simmons et al., in prep). 
The median $V-z$ color difference ($|{\rm [V-z]_{total}} -  {\rm [V-z]_{hostgalaxy}}|$) is 0.03 magnitudes, and two outlying sources with large color differences have low X-ray hardness ratios, HR $\sim-0.5$, consistent with unobscured AGN.
Therefore, for the moderate luminosity obscured AGN in our HST sample, the integrated galaxy color does not appear to be significantly affected by the central AGN light.

To further test the effects of the integrated AGN light, we combine synthetic stellar population models with unobscured quasar templates \citep{vandenberk2001,richardsetal2006}, creating model SEDs containing the post-starburst template used in Figure \ref{fig:col} and up to 50\% of light (normalized in the R-band) from the quasar template.
We then convolve these  composite spectra with our filter set  and use EAzY to determine the rest frame colors.  
The $U-V$ colors are preferentially shifted towards the blue, but the net effect is less than 0.1 mag (and typically $<0.03$ mag).
However, we do note that the very blue, very luminous AGN (Figure \ref{fig:cmass}, bottom right) have high X-ray luminosities and hardness ratios consistent with unobscured AGN.  
In these SEDs, the blue continuum is very flat and the AGN light dominates the integrated light, causing  the blue colors at optical wavelengths.
\citet{pierceetal2010} also simulate the addition of unobscured QSO light to template galaxy SEDs (contributing up to 50\% of the B-band flux), finding that the additional blue continuum from the QSO can bias the $U-B$ colors by up to a magnitude.
For the obscured AGN,  the light from the central AGN  continuum distribution does not greatly affect the AGN host galaxy colors.

The addition of the unobscured quasar light to a young stellar population makes the shorter wavelength colors slightly blue, but at longer wavelengths the composite is redder (${\rm [V-J]_{AGN+gal}\sim [V-J]_{gal} + 0.2}$).
However, for models several hundred million years post starburst (i.e., at redder $U-V$ colors), the AGN light does not change the rest-frame $V-J$ color enough to move it into the dusty region of the diagram (${\rm [V-J]_{AGN+gal}\sim [V-J]_{gal} + 0.02}$).
Therefore, our finding that $\sim$50\% of the AGN host galaxies live in passively evolving host galaxies is robust to contamination from a central AGN source.

\begin{figure*}[ht]
\includegraphics[angle=0, width=0.97\textwidth]{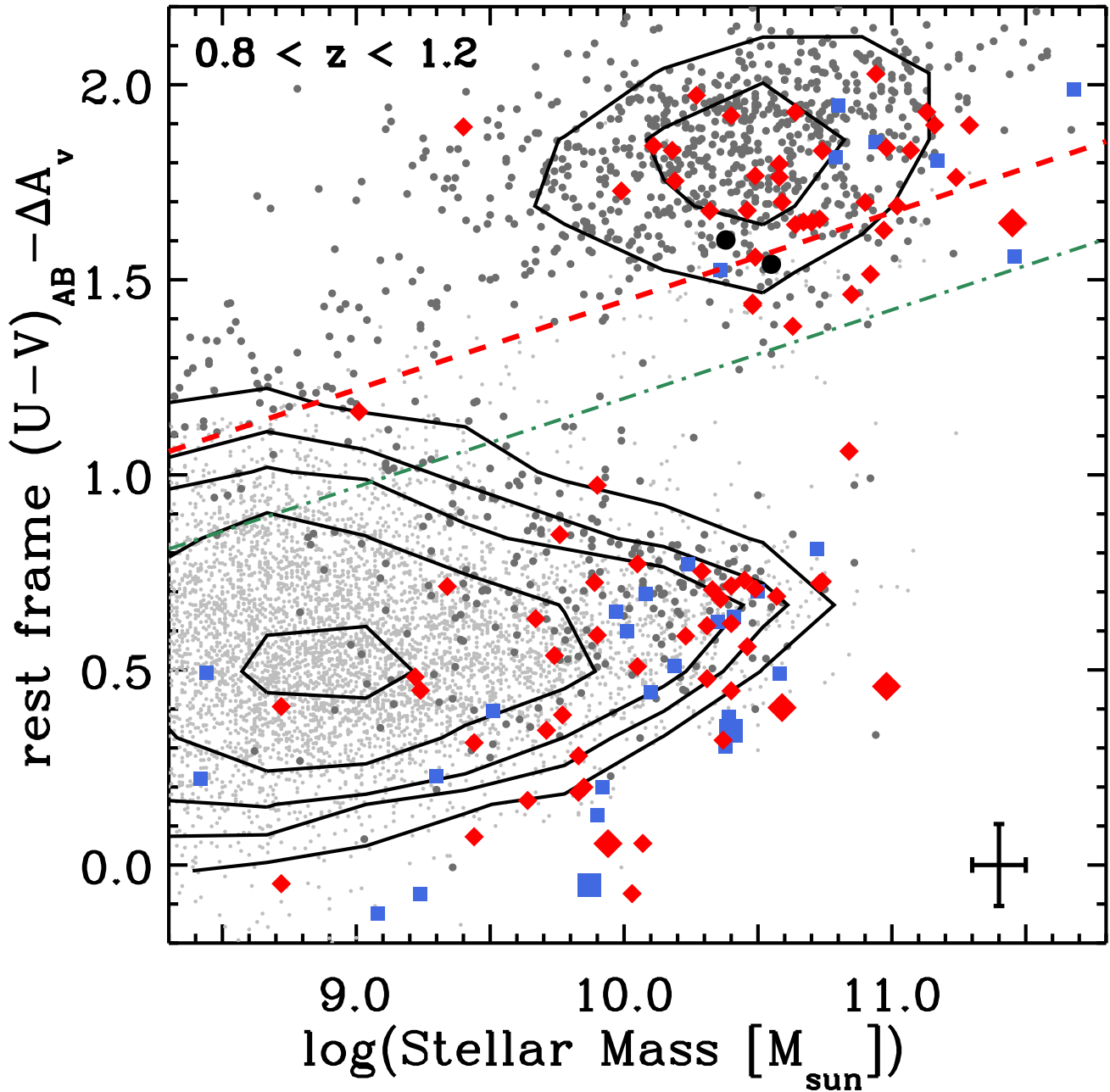}
\caption{The dust-corrected color-mass diagram shows that AGN and galaxies divide cleanly into red and blue colors  (points as in Figure \ref{fig:cmass}).  Half of AGN host galaxies at $z\sim1$ remain on a passive red sequence, while the other half lie at the massive end of the blue cloud.
The location of the AGN host galaxies on the blue edge of the red sequence suggests that these AGN may be recent arrivals to the red sequence which will redden further as their stellar populations age.}
\label{fig:massav}
\end{figure*}

Following \citet{brammeretal2009}, we de-redden the rest-frame colors based on the $A_V$ from the SED fit assuming a \citet{calzettietal2000} dust law ($\Delta A_V = 0.47 \times A_V$).
Figure \ref{fig:massav} shows that many galaxies in the green valley are removed in the dust-corrected $U-V$ color diagram.
That is, the population of AGN host galaxies divides into blue and red populations after correcting the $U-V$ colors for the contribution of dust, just as has been seen for non-active high redshift galaxies  \citep{cowiebarger2008,brammeretal2009}.
Roughly half the AGN remain on the lower edge of the red sequence, while the other half has  blue de-reddened colors.
These corrected colors depend on accurate determinations of $A_V$ from the SED fitting, which can be overestimated in galaxies that experienced a modest starburst several hundred million years ago.  
However, 95\% of sources fitted by FAST with $A_V > 0.2$ fall outside the passive selection region in Figure \ref{fig:col}, suggesting overestimated $A_V$ values may affect at most 5\% of the population.  
Follow-up deep spectroscopy is underway to confirm these AGN host galaxies' dusty-star-forming nature.

\section{Discussion \& Conclusions}

Using rest-frame near-infrared colors, we have shown that few AGN host galaxies in the green valley at $z\sim1$ are genuinely transitioning from blue to red; 
75\% of the host galaxies with green observed colors are intrinsically blue but reddened by dust.
Corrected for reddening, the observed $z\sim1$ AGN population is split evenly between galaxies in the blue cloud and those on the red sequence.

Some models of galaxy evolution suggest that heating of the interstellar medium, due to feedback from black hole growth, could potentially quench star formation \citep{silkrees1998,faberetal2007,schawinskietal2007,dimatteoetal2008}. 
The dusty, intrinsically blue, star-forming AGN host galaxies are fully consistent with the observed 
AGN heating up the interstellar gas and shutting down star formation.
In that case, the galaxies with dust corrected intermediate colors, $\sim5$ green valley AGN hosts and $\sim$50 non-AGN, would be transitioning from the blue cloud.  
Since there are $\sim$40 AGN with dust-corrected blue star forming hosts, this implies that these galaxies spend comparable times in a star-forming phase and passively evolving green valley phase.
However, this is not evidence that AGN quenching occurs.
Follow-up studies with deep optical and infrared spectroscopy can more accurately age-date the stellar populations and explore this possible link.

The red AGN host galaxies are actually the same color --- and thus are likely to have the same post-starburst age --- as comparable X-ray-selected samples of AGN at $z\sim0$ \citep{schawinskietal2009agn}. 
They lie on the red sequence at $z\sim1$ but in the green valley at $z\sim0$ due to the evolution of the red sequence from z$\sim0-1$ \citep[e.g.,][]{belletal2004}.
Even with an effectively instantaneous cessation of star formation, a starburst takes $\sim300$ Myrs to age to colors consistent with the bluest edge of the $z\sim1$ red sequence, and more gradual exponential declines of star formation take $\sim 500$ Myrs or more.
Given a typical AGN lifetime, ($\sim10-100$ Myrs; \citealt{ferrareseford2005}), the redder host galaxies are slightly older than one would expect if the star formation was shut down part way through the current episode of AGN activity.


Furthermore, if the presently observed AGN is responsible for the cessation of star formation, then the relative frequency of dust-corrected red AGN hosts compared to dust-corrected green AGN hosts gives a rough estimate of the duty cycle of AGN activity.
That is, because there are at least 10 times as many AGN on the red sequence as in the green valley (for log$M_{star} \ge 10 M_\odot$, where our sample is complete), a continuously active AGN would have to spend 10 times as long on the red sequence as their stellar populations take to age through the green valley.
Given the ages of post-starburst galaxies with colors of $U-V\sim1.5$, this implies an AGN lifetime over $\sim1$ Gyr.
It therefore seems more likely that  AGN activity is episodic, with the current episode of accretion beginning after the star formation was turned off.  
In either case, the AGN in red galaxies could represent a ``maintenance mode," in which enough energy is released into the interstellar medium to prevent future episodes of star formation \citep{crotonetal2006,hopkinsetal2006}.
It is interesting to note that the lowest luminosity AGN in our sample, ${\rm log (L_{XR}) \sim 42-42.5 \rm{~ergs~s^{-1}}}$, are preferentially red and passive. 
That is, a lower level of AGN activity may suffice to prevent gas cooling, whereas a more luminous AGN might be required to quench vigorous star formation.

The two populations of AGN hosts at $z\sim1$ --- in young star-forming galaxies and in older, passively evolving galaxies --- appear to represent distinct phases of AGN activity.
Specifically, AGN in young, dusty star-forming galaxies are prime candidates for quenching star formation and driving their host galaxies towards the red sequence, while AGN in much older, red host galaxies may be may be keeping the interstellar gas hot and preventing future episodes of star formation.

\section{Acknowledgements}
We thank the referee for helpful comments which improved this paper.
Support from NSF grants AST-0407295, AST-0449678, AST-0807570
and Yale University is gratefully acknowledged.
Support for the work of ET and KS was provided by  NASA through Einstein
Post-doctoral Fellowship Award Numbers PF8-90055 and PF9-00069
issued by the Chandra X-ray Observatory Center, which is
operated by the Smithsonian Astrophysical Observatory for and on
behalf of the National Aeronautics Space Administration under contract
NAS8-03060.

\end{document}